\documentclass{article}

\usepackage{amsmath}
\usepackage[utf8]{inputenc}
\usepackage{gensymb}
\usepackage{graphicx}
\usepackage{dblfloatfix}
\usepackage{authblk}
\usepackage{color}

\usepackage{geometry}
\DeclareMathOperator{\sinc}{sinc}

\title{High magnetic field stability in a planar graphene-NbSe$_2$ SQUID} 

\author[1,2]{Ayelet Zalic} 
\author[3]{Takashi Taniguchi}
\author[4]{Kenji Watanabe}
\author[1,5]{Snir Gazit}
\author[1,2]{Hadar Steinberg}
\affil[1]{The Racah Institute of Physics, The Hebrew University of Jerusalem, Jerusalem 91904, Israel}
\affil[2]{The Center for Nanoscience and Nanotechnology, Hebrew University of Jerusalem, Jerusalem 91904, Israel}
\affil[3]{International Center for Materials Nanoarchitectonics, National Institute for Materials Science, 1-1 Namiki, Tsukuba 305-0044, Japan}
\affil[4]{Research Center for Functional Materials, National Institute for Materials Science, 1-1 Namiki, Tsukuba 305-0044, Japan}
\affil[5]{The Fritz Haber Research Center for Molecular Dynamics, The Hebrew University of Jerusalem, Jerusalem 91904, Israe}

\date{}

\begin{document}

\maketitle

\begin{abstract}
\textbf{ Thin NbSe$_{2}$ retains superconductivity at high in-plane magnetic field up to 30 T. In this work we construct an atomically thin, all van der Waals SQUID, in which current flows between NbSe$_{2}$ contacts through two parallel graphene weak links. This fully planar device remains uniquely stable at high in-plane field. This enables tracing the evolution of the critical current interference patterns as a function of the field up to 4.5 T, allowing  nm-scale sensitivity to deviation from a perfect atomic plane. We present numerical methods to retrieve asymmetric current distributions $J_0$(x) from measured interference maps, and suggest a new application of the dual junction geometry to probe the current density in the absence of phase information. The interference maps exhibit a striking field-driven transition, indicating a redistribution of supercurrent to narrow channels. Our results suggest the existence of a preferred conductance channel with enhanced stability to in-plane magnetic field. }
\end{abstract}

Transition metal dichalcogenide (TMD) superconductors such as NbSe$_{2}$ can be mechanically exfoliated to yield thin layers down to the monolayer limit ~\cite{Xi_2016, Tsen2016}. Thin NbSe$_{2}$ superconducting electrodes sustain very high in-plane magnetic fields ($B_\parallel$) beyond the Pauli limit, due to suppressed orbital depairing and Ising spin orbit coupling (ISOC) which locks spins in the out of plane orientation~\cite{Xi_2016}. The superconducting gap persists nearly unchanged up to 10 T~\cite{Dvir_Nat_Comm_2018}, and remains observable up to 25 T in tunneling measurements~\cite{DvirArxiv}.

\begin{figure}[ht!]
    \centering
    \includegraphics[width=0.9\textwidth]{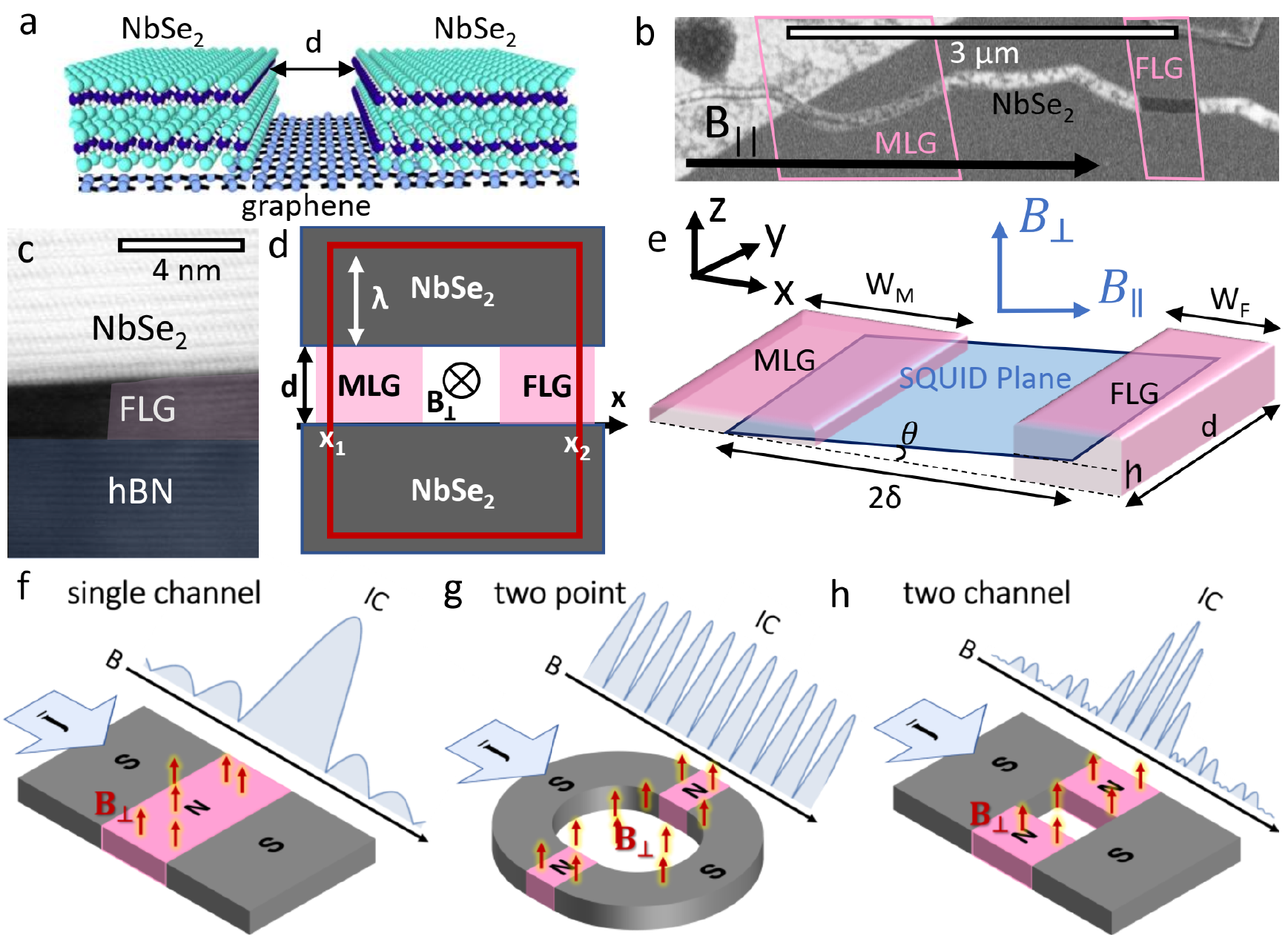}
    \caption{   \textbf{a.} Planar NbSe$_2$-graphene-NbSe$_{2}$ JJ geometry  \textbf{b.} SEM image of the device, flake outlines are a guide to the eye \textbf{c.} False-color cross-section TEM measurement of FLG region showing atomically clean interfaces between NbSe$_{2}$ and FLG, and between FLG and hBN \textbf{d.} Schematic showing $B_\perp$ flux through one possible current circulation path, with an area (2$\lambda$+d)$|x_2-x_1|$ \textbf{e.} Schematic illustration of FLG and MLG parallel weak links of different thicknesses. Directions $\hat{x}$ and $\hat{y}$ are in the plane of the flakes, $\hat{z}$ is perpendicular. Mean SQUID plane is shown in blue, at an angle $\theta$. $B_\parallel$ is parallel to the SQUID plane and $B_\perp$ is perpendicular to it. Crack length $d$ is in the direction of current flow \textbf{f.-h.} Illustration of interference patterns for different junction geometries: \textbf{f.} single channel (Fraunhofer), \textbf{g.} two point (SQUID), and \textbf{h.} two channel (Fraunhofer envelope modulates SQUID oscillation). }
    \label{fig:device}
\end{figure}

It is useful to incorporate thin TMD superconductors in devices which utilize their unique properties at high $B_\parallel$. NbSe$_{2}$ has been coupled laterally to graphene to realize NS junctions ~\cite{Efetov_2016,Yabuki2020}. Devices consisting of NbSe$_{2}$ flakes coupled on both sides of a narrow graphene channel (Fig.~\ref{fig:device}a) are well-behaved Josephson junctions (JJs)~\cite{Lee2019,Zalic_2021}. Our two dimensional planar Josephson junctions (2DJJs), constructed exclusively from van der Waals (vdW) materials by transferring a cracked NbSe$_{2}$ flake on top of a graphene flake, are unique in retaining a Josephson effect at high parallel magnetic fields.

The Josephson effect occurs when supercurrent flows between two superconducting electrodes (in this case NbSe$_2$) connected by a weak link (graphene). Upon application of a small (mT scale for our devices) magnetic field perpendicular to the junction ($B_\perp$), the superconducting order parameter $\Delta e^{i\varphi}$ acquires a position-dependent phase and undergoes interference. This leads to a Fourier relation between the critical current $I_C$($B_\perp$) and the maximal local critical current density $J_0(x)$ \cite{Tinkham}:

\begin{equation}
    \begin{aligned}
    I_C(B_\perp) = \bigg{|}\int_{-\infty}^{\infty} J_0(x)e^{ikx} dx\bigg{|} 
    \end{aligned}
\end{equation}

where $k\equiv\frac{2\pi(2\lambda+d)B_\perp}{\phi_0}$, such that a loop connecting any two points $x_1,x_2$, and extending across the junction length $d$ into the superconductors up to the London penetration depth $\lambda$, encloses a magnetic flux of $k(x_2-x_1)/2\pi$ in units of $\phi_0$ (see Fig.~\ref{fig:device}d). In our previous work, we studied a 2DJJ subject to parallel fields as high as 8.5 T~\cite{Zalic_2021}. Even at such high fields, the highest accessible to our magnet, the device exhibits interference patterns.

The $I_C(B_\perp)$ interference pattern holds information about the current density distribution within the junction, according to Eq. 1. However, extracting this information is difficult, since interference patterns reflect a variety of physical effects and require appropriate assumptions and methods to reconstruct the lost phase. Using such reconstructions, several groups have shown that current in graphene based Josephson junctions flows through the bulk with enhanced current density along the edges, while at the Dirac point edge channels dominate junction transport \cite{Allen2016,Zhu2017}.

In this work we extend the all-vdW 2DJJ concept to a SQUID geometry, with current flowing between NbSe$_{2}$ contacts through parallel monolayer graphene (MLG) and few-layer graphene (FLG) weak links (see Fig.~\ref{fig:device}b). 
In this highly planar structure, the graphene flakes are supported by a flat, insulating hexagonal boron nitride (hBN) substrate and all interfaces are atomically clean (see Fig.~\ref{fig:device}c).
The device is gate tunable via a SiO$_{2}$ back-gate with traceable $I_C$($B_\perp$) interference patterns up to $B_\parallel = $ 4.5 T. The stability of the device allows us to extract the spatially varying current density as a function of gate voltage $V_G$ and $B_\parallel$, revealing signatures of device geometry including a  nm-scale step height between the MLG and FLG planes (see Fig.~\ref{fig:device}e). 

At high $B_\parallel$ the Fraunhofer envelope due to the current distribution within the MLG disappears, indicating a narrowing of the current channel. In order to extract the current density distribution, we address the phase retrieval problem using both a minimal parameter analytical fit, and a constrained maximum entropy fitting procedure. Further, we develop a new method relying on the Wiener Khinchin theorem, where a narrow junction serves to map the current density of a wider junction. Using these methods, we observe a gate-dependent spatial variation of current density in the MLG weak link at low field; at high $B_\parallel$ our analysis corroborates the observed disappearance of the Fraunhofer envelope, confirming that current in the MLG focuses into a single narrow channel. 

For our SQUID we use a single cracked NbSe$_{2}$ flake, of approximately 13  nm thickness, as seen in the cross-section TEM measurement (see supplementary); the length of both junctions, imposed by the NbSe$_{2}$ crack, is $d$ = 140 nm in the direction of the current flow (see Fig.~\ref{fig:device}d). We measure dimensions using SEM, see Fig.~\ref{fig:device}b: the width of the MLG flake junction is $W_M\approx1.45$ $\mu$m, while the width of the FLG junction is $W_F\approx 0.45$ $\mu$m. The distance between the centers of the two junctions is $2\delta\approx2.7$ $\mu$m. For what follows, it is important to distinguish between different planes of reference in the sample. The magnetic field $B_\parallel$ we refer to as ``in-plane" is oriented parallel to the mean SQUID plane: the plane connecting the center of the MLG and FLG flakes (see Fig.~\ref{fig:device}e). This plane is at a small angle $\theta$ with respect to the plane of the MLG flake. 
Throughout the measurements reported here, $B_\parallel$ is kept strictly aligned to the SQUID plane, for minimizing flux jumps. 
$B_\perp$ is defined geometrically perpendicular to $B_\parallel$. Note that we control the field along the axes of the lab magnets, which are not exactly aligned with the SQUID plane; we describe the compensation and alignment procedure in detail in the supplementary.

To gain initial insight into the expected $I_C$($B_\perp$) in the SQUID, we make two simplifying assumptions: (i) the phase dynamics are local and (ii) the current-phase relation is sinusoidal. Both assumptions are typical for graphene-based JJs. However, ballistic graphene JJs may exhibit measurable skewness in the current phase relation \cite{Nanda2017}, while
in ultrathin superconducting contacts, the Pearl length $\Lambda$ = $2\lambda_L^2/t$ replaces $\lambda_L$ as the relevant field decay scale and the dynamics become potentially non-local \cite{Rodan-Legrain2021}. 
We also approximate a spatially uniform current density in each channel. Applying Eq. 1 to this simplified model produces a two channel diffraction pattern \cite{Jaklevic1965} (see Fig.~\ref{fig:device}h), with the finite FLG/MLG widths ($W_F,W_M$) generating Fraunhofer-like envelopes (Fig.~\ref{fig:device}f) modulating the SQUID oscillations (Fig.~\ref{fig:device}g). 
The angle of the SQUID plane with respect to the MLG and FLG plane directly translates into a phase difference between the two channels. We describe the exact expression and the calculation leading to it in the supplementary material. Below, we refer to this as the ``Analytical Model''.

\begin{figure}[ht!]
    \centering
    \includegraphics[width=0.9\textwidth]{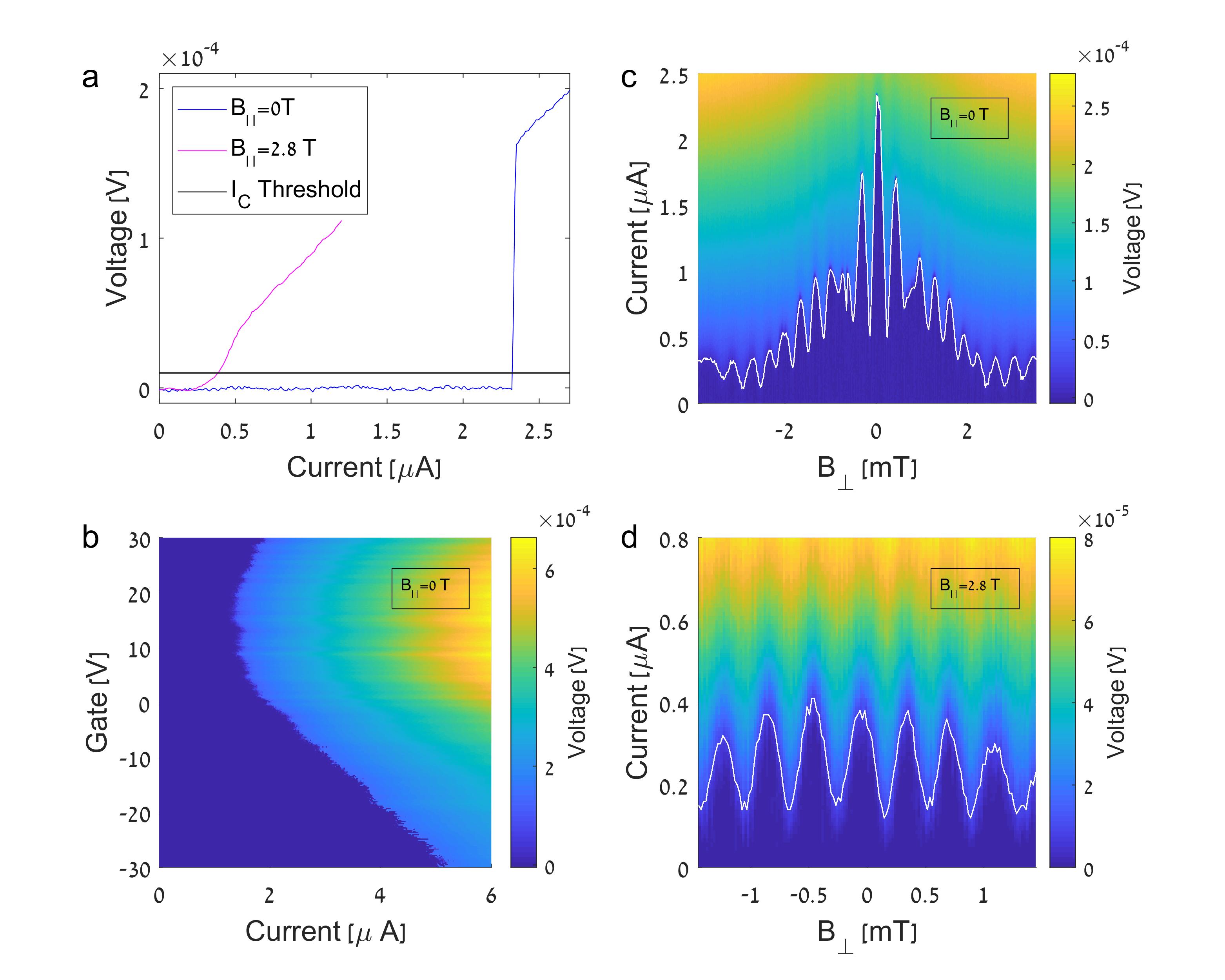}
    \caption{  \textbf{a.} Current-voltage traces at $B_\parallel$ = 0 T, $B_\parallel$ = 2.8 T. Voltage threshold for determining critical current shown in black \textbf{b.} Current-voltage traces as a function of gate voltage at zero field. Gate modulates the critical current, with a minimum around 10 V (the MLG Dirac point) ~\textbf{c.} Interference pattern at $B_\parallel$ = 0 T, with a $\Delta B \approx$ 380 $\mu$T oscillation corresponding to the SQUID area, and an envelope reflecting the area of the MLG and FLG junctions. A white line marks the threshold detection of $I_C$ \textbf{d.} Interference pattern at $B_\parallel$ = 2.8 T. The SQUID oscillation maintains similar periodicity to (a), but the envelope is no longer visible. Note measured $B_\perp$ range at $B_\parallel$ = 2.8 T is smaller than at 0 T to avoid entry of vortices }
    \label{fig:meas1}
\end{figure}

We begin by showcasing the basic properties of the 2D SQUID in Fig.~\ref{fig:meas1}. Current-voltage characteristics of the SQUID switch from zero to finite resistance at the critical current $I_C$, which we define according to a voltage threshold (see Fig.~\ref{fig:meas1}a). The transition from superconducting to normal conductance is sharpest at $B_\parallel$ = 0 T. Panel (b) illustrates the modulation of critical current by varying charge carrier density. In our SQUID the common back-gate tunes both FLG and MLG densities simultaneously, therefore it is not possible to pinpoint the MLG Dirac point exactly (it is likely in the region of minimal $I_C$ around 10 V). 

The interference pattern of $I_C$($B_\perp$) at zero gate voltage and $B_\parallel$ = 0 T is shown in Fig.~\ref{fig:meas1}c. The rapid oscillations of $I_C$, with a magnetic field period $\Delta B$ = $\phi_0 A_{SQ}$ $\approx$ 380 $\mu$T, reflect the area of the SQUID $A_{SQ}$ = $2\delta(2\lambda + d)$ = 5.4 $\mu$m$^2$. This area implies an effective penetration length $\lambda$ = 930 nm, longer than $\lambda_L$ $\approx$ 200 nm yet shorter than the Pearl length $\Lambda$ = $2\lambda_L^2/t\approx$ 6 $\mu$m. 
The SQUID oscillations are modulated by an envelope which derives from the areas of the MLG and FLG channels, as illustrated by a schematic of two channel interference in Fig.~\ref{fig:device}h. Note that the measured $B_\parallel$ = 0 T pattern is not perfectly symmetric with respect to $B_\perp$. This could be a signature of various symmetry breaking effects \cite{Rasmussen_2016, Assouline_2019}, but is most likely due to vortices in the vicinity of the junction or a small trapped parallel flux. 

The introduction of $B_\parallel$ dramatically changes this pattern. At $B_\parallel$ = 2.8 T (panel (d)), the SQUID oscillation persists and maintains its periodicity, whereas the envelope is no longer discernible. The data now resembles the two-point interference pattern in Fig.~\ref{fig:device}g. 
The contrast between zero and high field interference patterns is one of the main new results of our work. The transition to a two-point SQUID indicates a change in the supercurrent distribution, which becomes focused within a narrow channel at higher fields.

\begin{figure}[ht!]
    \centering
    \includegraphics[width=0.9\textwidth]{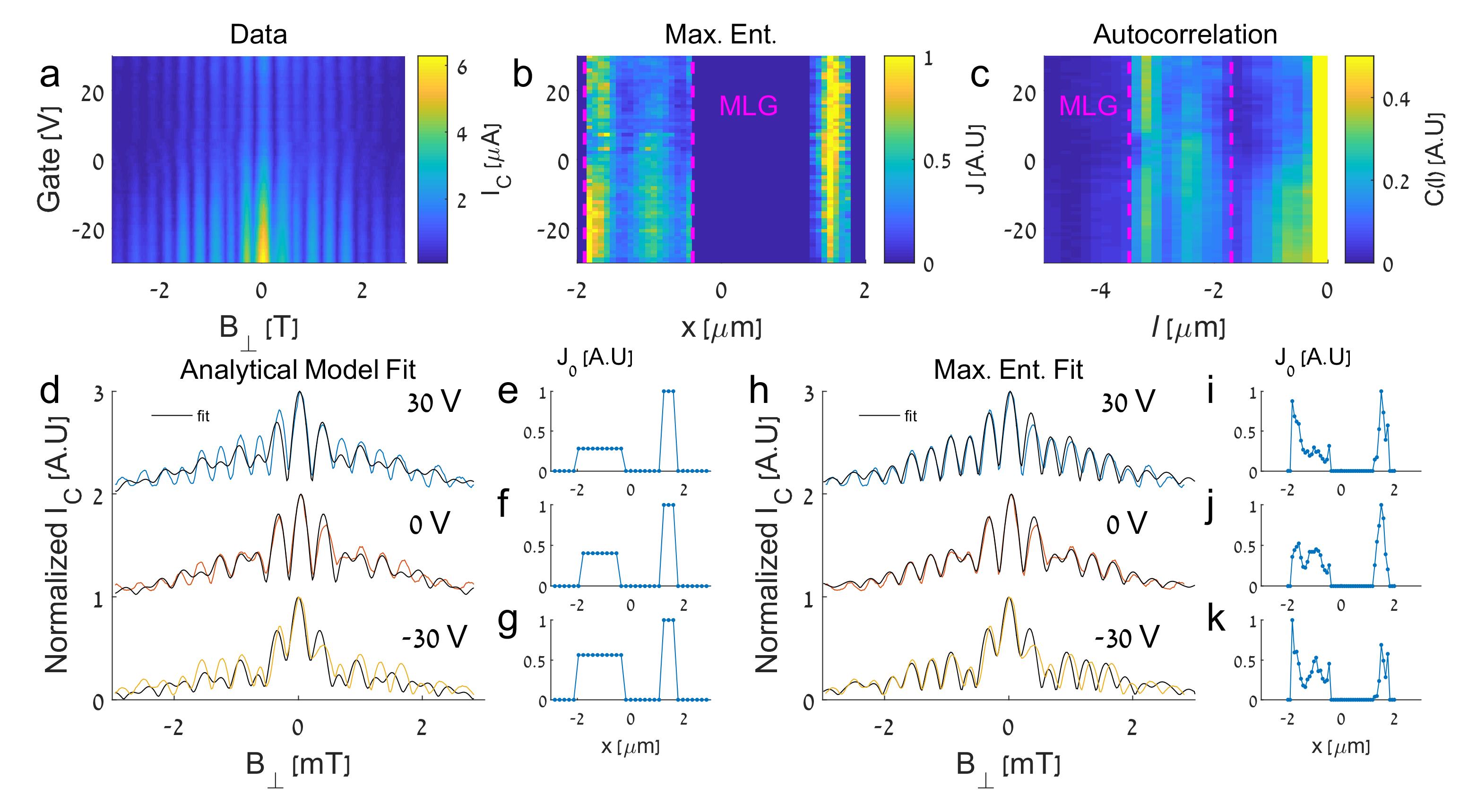}
    \caption{   ~\textbf{a.} $I_C$ (color scale) vs. $B_\perp$ and gate voltage ~\textbf{b.} Normalized current density extracted from panel (a) using maximum entropy method ~\textbf{c.} Normalized autocorrelation function of the current density (color scale). Compare the side band to the MLG current density in panel (b) ~\textbf{d.} Interference patterns at gate voltages of -30, 0, 30 V with two channel analytical fit  \textbf{e.-g.} Current density profiles corresponding to the analytical fits in panel (d)  ~\textbf{h.} Interference patterns at gate voltages of -30, 0, 30 V with maximum entropy fit  \textbf{i.-k.} Current density profiles corresponding to fits in panel (h) }
    \label{fig:gate}
\end{figure}

To investigate the supercurrent distribution systematically, we first turn to study how the $B_\parallel$ = 0 T interference pattern evolves with respect to applied gate voltage, similarly to earlier works \cite{Allen2016,Zhu2017}. We measure $I_C$ as a function of $B_\perp$ and $V_G$ continuously, as shown in the color-plot in Fig.~\ref{fig:gate}a. The data indicate that the overall SQUID periodicity remains fairly constant, whereas the critical current magnitude and the envelope change - indicating a variation in current distribution. Fig.~\ref{fig:gate}d shows selected interference patterns from (a) at $V_G$ = -30 V, 0 V, 30 V.
We fit these curves using the analytical model described above, where the free parameters are the MLG and FLG widths $W_M,W_F$, the distance between their centers $2\delta$ (see Fig.~\ref{fig:device}e), and the ratio between their critical current densities $J_M/J_F$ (see supplementary for details). 

Fig.~\ref{fig:gate}e-g shows the current density profile for each $V_G$ trace, extracted from the fit. For $2\lambda + d$ = 2 $\mu$m, we find that fit parameters $W_M,W_F,\delta$ agree with the dimensions determined from SEM measurements shown in Fig.~\ref{fig:device}d (see comparison table in the supplementary). The model also indicates that the graphene channel exhibits a stronger response to the gate. The analytical curves fit the data reasonably well for the central lobes of the interference patterns - but the higher order lobes are far more pronounced in the data compared with the fit, hinting that the current density distribution has finer spatial detail beyond the two uniform conduction channels.

We thus turn to extract the current distribution in greater detail. Since the interference pattern reflects the absolute value of the Fourier transform of the current density, phase information is lost and it is impossible to directly apply an inverse Fourier transform. The oft-cited Dynes-Fulton approach to phase retrieval assumes a nearly symmetric current distribution, and so is not applicable in our case \cite{Dynes_1971}. This approach is also not self-consistent, i.e. interference patterns generated from the extracted current distribution may not fit the original data \cite{Hui_2014}. 

An alternative method postulates a current density profile sampled at $N$ discrete spatial points and subject to known physical constraints to calculate the critical current via a forward Fourier transform \cite{Hui_2014,Ghatak2018}. The current density profile is then adjusted to obtain the best fit of the calculated interference pattern to the data. One work uses a constrained nonlinear optimization algorithm to minimize the least squares difference of the data and the fit \cite{Hui_2014}; another work adds a maximum entropy constraint in order to avoid spurious sharp changes in current density \cite{Ghatak2018}. We use an approach inspired by the latter, which we term ``the maximum entropy method" for future reference. See supplementary for the full details of our fitting algorithm, including our approach  towards calibrating parameters and avoiding over-fitting (based on the L curve \cite{Hansen_1992}).  

Using the maximum entropy method, we extract the current density at $V_G$ = -30 V, 0 V and 30 V, shown (normalized by the maximal $J_0$ for each gate) in Fig.~\ref{fig:gate}i-k. 
To confirm self-consistency we apply Eq. 1 to reproduce the interference pattern corresponding to the extracted current density. We compare these to the measured patterns in Fig.~\ref{fig:gate}h; the obtained fit is indeed far better than the analytical fit in Fig.~\ref{fig:gate}d, especially in the higher order lobes.
In panels i-k, the current within the MLG has two peaks; perhaps these are the edge channels observed in references \cite{Allen2016,Zhu2017}. The color plot in Fig.~\ref{fig:gate}b shows the full evolution of the extracted current density with gate voltage. 

The maximum entropy fit is complex, with many algorithmic as well as physical parameters. This leads us to introduce a new method which uses the narrower FLG junction as a direct probe of the current density in the wider MLG junction. This method leans on the Wiener Khinchin theorem, which states that the energy spectral density of a function and its autocorrelation $C(l)$ are Fourier transform pairs. In our case, $|I_C(B_\perp)|^2$ is the energy spectral density of $J_0(x)$, and thus:

\begin{equation}
        F(|I_C(B_\perp)|^2) = C(l) = \int_{-\infty}^{\infty}J_0^*(x) J_0(x+l) dx
\end{equation}

Consider a two channel device, where the current density in the one channel is narrow, approximated by the Dirac delta function, whereas the current in the other channel is widely distributed. The separation between the centers of the two channels is larger than their combined widths. The autocorrelation of the current density in this case contains a term equal to the current density in the wider channel (see calculation in the supplementary). In our case the FLG is only a few times narrower than the MLG, and carries a similar total current. In this instance, the autocorrelation convolves the FLG and MLG densities, resulting in a feature which qualitatively resembles the MLG current density ``smeared" at the scale of the FLG width, and centered at $l$ = $-2\delta$ equal to the distance between the centers of the two channels. Fig.~\ref{fig:gate}c shows the autocorrelation \footnote{The autocorrelation is smoothed as a result of zero-padding before calculating the discrete Fourier transform of $|I_C(B_\perp)|^2$.} as a function of gate; the resulting ``side-band" centered at $l$ = $-2.7$ $\mu$m is qualitatively similar in its form to the extracted current density in Fig.~\ref{fig:gate}b. 
We thus confirm that the autocorrelation method can be used to extract the current distribution, when one channel is narrow.

 \begin{figure}[h]
    \centering
    \includegraphics[width=0.9\textwidth]{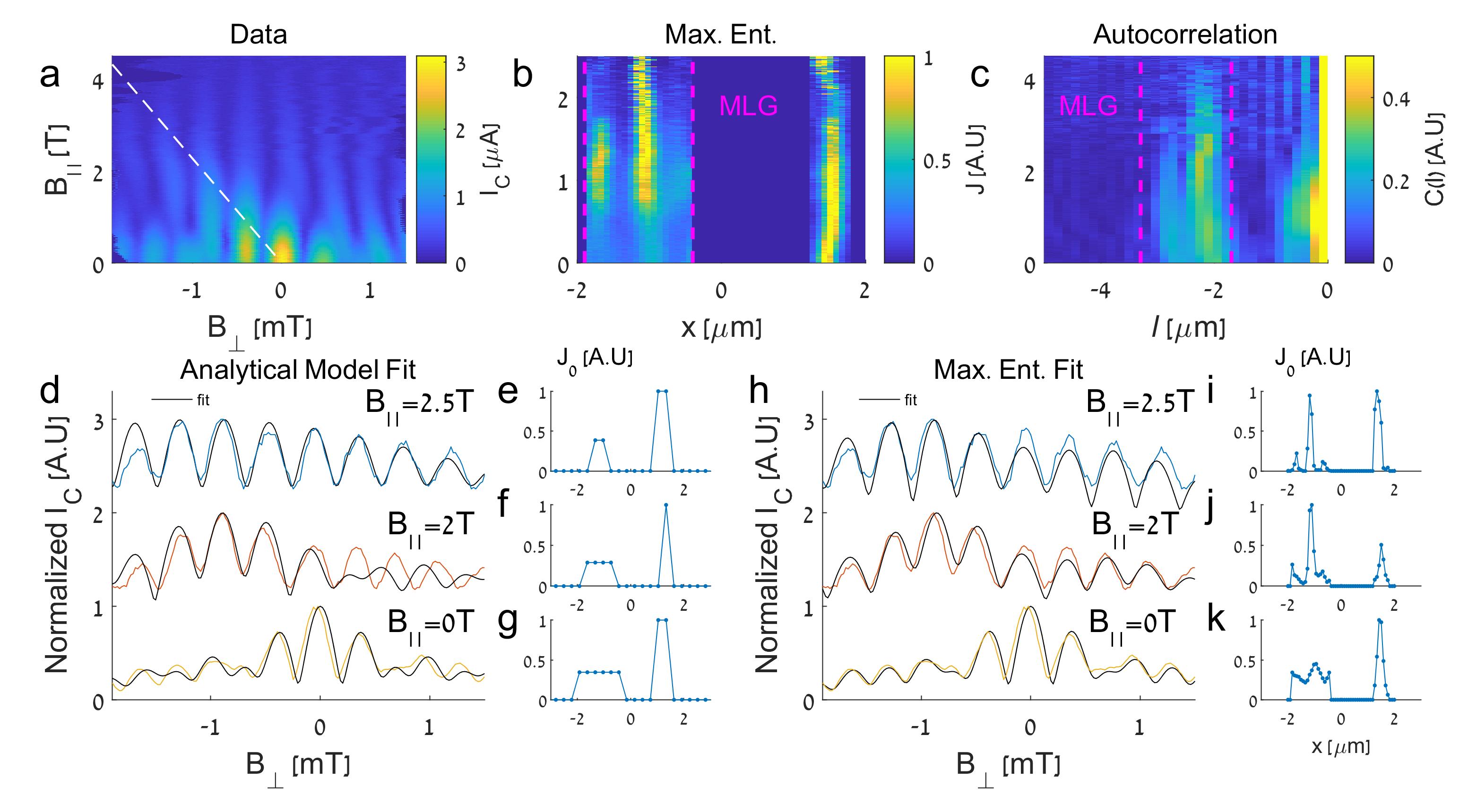}
    \caption{    ~\textbf{a.} $I_C$ (color scale) vs. $B_\perp$ and $B_\parallel$. White dashed line marks the geometric angle $\theta\approx0.025\degree$ between the SQUID plane and the MLG ~\textbf{b.} Normalized current density extracted using the maximum entropy method ~\textbf{c.} Normalized autocorrelation function of the current density (color scale). Compare the side band to the MLG current density in panel (b) ~\textbf{d.} Interference patterns at $B_\parallel$ 0 T, 2 T, 2.5 T with two channel analytical fit  \textbf{e.-g.} Current density profiles corresponding to the analytical fits in panel (d)  ~\textbf{h.} Interference patterns at $B_\parallel$ 0 T, 2 T, 2.5 T with maximum entropy fit  \textbf{i.-k.} Current density profiles corresponding to fits in panel (h). }
    \label{fig:field}
\end{figure}


We now return to the field-driven transition shown in Fig.~\ref{fig:meas1}, using the methods discussed above to reconstruct the current distribution.
Fig.~\ref{fig:field} follows the same structure as Fig.~\ref{fig:gate}. In panel (a) we plot $I_C(B_\perp,B_\parallel)$, tracing the evolution of the interference patterns as a function of parallel magnetic field at $V_G$ = 0 V. 
This interference plot is extremely stable, up to $B_\parallel$ = 4 T, barring minor flux jumps around $B_\parallel$ = 2.8 T, likely due to vortices entering the vicinity of the junction. 
There are two notable features seen in this data. First, the data exhibits a diagonal drift of the MLG envelope towards negative $B_\perp$. This is evident in a shift of maximal $I_C$, and of the first Fraunhofer nodes seen to intersect with the SQUID features.
This shift is a direct result of device geometry: the step height $h$ between MLG and FLG planes creates an angle $\theta$ between the SQUID plane and the MLG. 
With $B_\parallel$ aligned to the SQUID plane, the flux through the MLG and FLG flakes receives a $B_\parallel$ component, leading to the drift of the envelope (see full calculation in the supplementary). 
We note that the angle $\theta\approx0.025\degree$ extracted from the fit indicates a step height around 1  nm, while the TEM measured FLG thickness is 2.4  nm; this probably indicates a distribution of current throughout the FLG, with the mean SQUID plane determined by the center of the FLG flake. There could also be an additional step or curvature in the hBN outside the range of the TEM.
The 2DJJ SQUID is thus an extremely sensitive tool for tracking deviations from atomic planar geometry.

Panel (b) shows current density extracted using the maximum entropy fit of an $I_C(B_\perp,B_\parallel)$ dataset.\footnote{The data corresponding to the fit in Fig.~\ref{fig:field} panel (b) is shown in supplementary Fig. S3; it is a different measurement containing a few more lobes in the interference pattern with respect to panel (a). This improves the spatial resolution of the extracted current density, but loses stability at lower $B_\parallel$.} 
The fit confirms that the current density in the MLG redirects into a narrow channel. 
The interference patterns produced by the maximum entropy procedure fit most of the measured patterns closely, as shown for selected values of $B_\parallel$ = 0 T, 2 T, 2.5 T in Fig.~\ref{fig:field}h. The extracted current densities at this succession of fields is depicted in Fig.~\ref{fig:field}i-k, illustrating the narrowing of the current carrying channel in the MLG as $B_\parallel$ increases. \footnote{As the current channel grows narrow the pre-defined geometric constraint of the wide channel provides too much freedom, leading to reduction in the quality of the maximum entropy fit }  

This phenomenology is apparent also in our other methods of analysis.
Panel (c) shows the autocorrelation of the current density. The side-band located around $l$ = 2.7 $\mu$m is qualitatively similar to the extracted MLG current density shown in Panel (b), and also exhibits a narrowing of the current channel in the MLG commencing at $B_\parallel$ = 2 T. Panel (d)  shows the best fit of our analytical model to interference patterns at $B_\parallel$ = 0 T, 2 T, 2.5 T, while Fig.~\ref{fig:field}e-g show the current density profiles corresponding to the fit. Here too, the current density in the MLG grows narrower as $B_\parallel$ increases. At low $B_\parallel$ the analytical curve does not fit the data of the higher order lobes; but as $B_\parallel$ increases the fit afforded by two uniform channels improves.

The transition towards narrow supercurrent channels has already been hinted at in our previous work - indeed, multiple diffusive MLG-NbSe$_{2}$ junctions also undergo a transition to SQUID-like interference patterns where all lobes are of similar height at high $B_\parallel$ \cite{Zalic_2021}.
In that work however, the patterns were too disordered to fit to equation 1, and we could not rule out the role of ripples due to the SiO$_2$ substrate~\cite{Zalic_2021}. 
In the present work, the device is flat due to the use of an hBN substrate. In addition, the signal is sufficiently stable to allow quantitative fitting. All models, assuming an experimental geometry corroborated by SEM and TEM, yield a clear transition between a distributed current density in the MLG at low fields, to a narrow supercurrent channel at high $B_\parallel$. 

We note that a similar effect of SQUID-like interference patterns at high $B_\parallel$, seen by Suominen et al., was attributed to suppression of supercurrent in the bulk of the JJ due to a magnetic dipole formed by tilted flux lines~\cite{Suominen2017}. In our geometry however, SQUID-like interference indicates one channel in the MLG, not necessarily on the edge. Moreover, the model in \cite{Suominen2017} is relevant for magnetic fields oriented parallel to the current flow whereas our $B_\parallel$ is perpendicular. The flux focusing effect is also weaker in thin NbSe$_{2}$ electrodes, where the tilt of the field lines is minimal due to a long London penetration length.

Thus in our devices, the origin of field-induced current re-distribution is an open question.
The SQUID-like interference indicates the presence of a conductance channel with superior resilience to high $B_\parallel$.
The suppression of a 2DJJ supercurrent vs. $B_\parallel$ is determined by the interplay of the Thouless and Zeeman energy scales~\cite{Buzdin2005,Zalic_2021}.
The Thouless energy is a transport energy scale, determined by the inverse of the traversal time of the junction \cite{Pannetier2000}. It supplants the superconducting gap as the determinant energy scale in a diffusive Josephson junction.
Hence, the superior resilience of a single channel could be the consequence of a higher Thouless energy, if a particular channel allows faster traversal of the junction. This could be, for example, a guided edge mode, or a shorter channel in a non-uniform junction geometry. However, the presence of a similar effect in a number of devices suggests that it is not related to a particular geometry.
Favored channels could also be the ones which experience minimal scattering in a disordered potential landscape.
Alternatively, graphene could inherit Ising spin-orbit coupling by proximity to the NbSe$_{2}$ within the extended contact region between the two materials \cite{Gmitra2015}. Such an interaction would enhance the stability of the carriers to in-plane field, and spatial variation of the induced coupling could lead to preferred channels. 
All in all, we find that the in-plane magnetic field appears to create narrow superconducting channels in graphene-NbSe$_{2}$ 2DJJs, an intriguing effect which has yet to be understood.

To conclude, the 2DJJ SQUID reported here is a novel experimental platform allowing us to measure interference patterns and extract current distributions at high $B_\parallel$, leading to interesting applications in sensing. 
It allows alignment of the parallel field to the SQUID plane at millidegree precision, and serves as an extremely sensitive probe for variation in the planarity of the current distribution in the weak links. 
Deviations from strict planar geometry may include substrate curvature and atomic height variation. 
Finally, we detect the presence of a supercurrent channel exceptionally stable to magnetic fields, raising important questions about the nature of the supercurrent in the graphene-NbSe$_2$ junction.

There is a recent surge of interest in planar JJs, realized in weak links with pronounced SO coupling. It is driven by the prospect of inducing 0-$\pi$ transitions~\cite{Li_2019, Ke_2019}, mapping spin textures using interference patterns~\cite{Hart_2016, Chen_2018}, and by predictions for topological effects tuned by parallel magnetic field~\cite{Pientka_2017,Ren_2019,Fornieri_2019, Mayer2019B}. Looking to the future, it will be interesting to explore proximity effects in the 2DJJ and 2D SQUID. Within the extended contact region between the graphene and NbSe$_{2}$ leads (see Fig.~\ref{fig:device}a), graphene could inherit Ising spin-orbit protection of the proximity superconducting gap \cite{Gmitra2015,Island2019,Gani2019}. 
Devices of this type would be useful in pinning down the role of the spin-orbit effect in the hybrid graphene-TMD structure.

\section{Acknowledgements}
The authors wish to thank Y. Anahory, M. Aprili, E. Grynszpan, N. Katz, A. Keselman, C. H. L. Quay and P. Ramachandran for insightful discussions. This work was funded by Israel Science Foundation Quantum Initiative grant 994/19, Israeli Science Foundation grant 861/19 and BSF grant 2016320. S.G. acknowledges support from the Israel Science Foundation, Grant No. 1686/18. A.Z. is grateful to the Azrieli Foundation for Azrieli Fellowships. K.W. and T.T. acknowledge support from JSPS KAKENHI (Grant Numbers 19H05790, 20H00354 and 21H05233).

\section{Author contributions}
A.Z fabricated the devices and performed the measurements, data analysis, analytical and numerical simulations. All authors contributed to the writing of the manuscript.

\section{Competing financial interests}
The authors declare no competing financial interests.

\section{Methods}
We exfoliate hbN on markered SiO$_2$ and locate substrate flakes of thickness around 20-40  nm. We exfoliate graphene to SiO$_{2}$ directly, and NbSe$_2$ first to PDMS and then stamp the PDMS on SiO$_{2}$ to transfer the flakes. This method supplies large, thin flakes of NbSe$_{2}$ which are not obtained by exfoliating directly from the blue tape to SiO$_{2}$. We use an optical microscope to search for two long, narrow graphene flakes which are within a few $\mu$m distance of each other for the channels of the SQUID, as well as NbSe$_{2}$ flakes which are a few layers thick and have an observable crack, less than 500  nm wide. We then employ a successive polycarbonate (PC) pickup technique ~\cite{Zomer2014} to pick up first the NbSe$_2$, then the graphene strips oriented perpendicular to the crack, and finally deposit the stack on the hBN substrate. We apply standard e-beam lithography and e-beam evaporation to create Ti/Au contacts to the NbSe$_2$, removing surface oxide using in-situ Argon ion milling prior to evaporation. Four-probe  measurements were conducted in a BluFors dilution cryostat with a 3T/9T vector magnet and base temperature of 20 mK.

\setcounter{equation}{0}
\setcounter{figure}{0}
\setcounter{table}{0}
\setcounter{page}{1}
\setcounter{section}{0}
\renewcommand{\theequation}{S\arabic{equation}}
\renewcommand\thesection{S\arabic{section}}
\renewcommand\thefigure{\textbf{S\arabic{figure}}}   
\renewcommand{\figurename}{\textbf{Supplementary Figure}}
\setcounter{secnumdepth}{1}

\section{Supplementary Section: Parallel field alignment procedure}


\begin{figure}[ht!]
    \centering
    \includegraphics[width=0.9\textwidth]{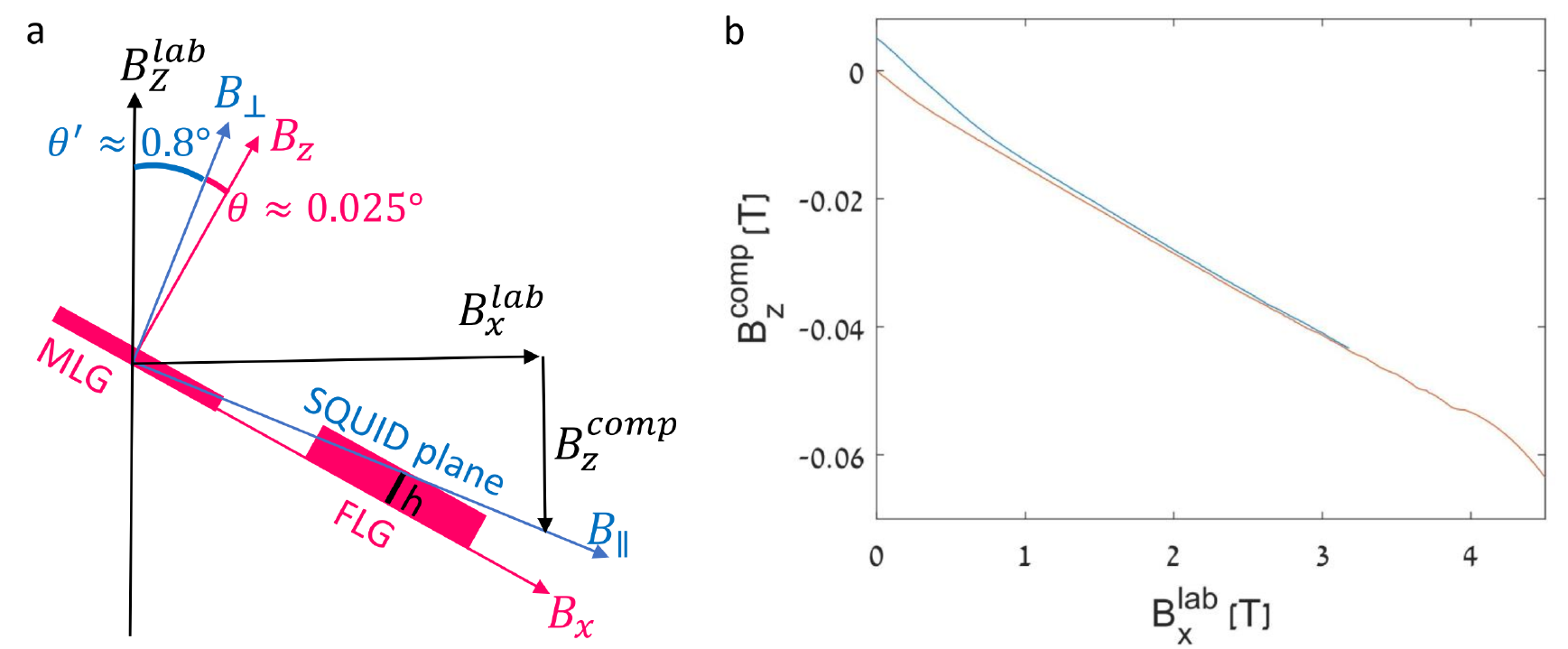}
    \caption{   ~\textbf{a.} Three axes systems: in black, the ``lab'' axes along which magnetic field is controlled. In blue, the SQUID plane, oriented at $\theta'=0.8\degree$ with respect to the lab frame. In pink, the MLG plane, oriented at $\theta=0.025\degree$ with respect to the SQUID plane. Applying $B_x^{lab}$ together with the compensation field $B_z^{comp}$ creates $B_\parallel$ in the SQUID plane   ~\textbf{b.} $B_z^{comp}$ as a function of $B_x^{lab}$ for the data presented in Fig.~\ref{fig:S2}c (blue) and main text Fig. 4 (orange)} \label{fig:S1}
\end{figure}


Our SQUID is highly sensitive to perpendicular fields on the scale of tens of $\mu$T. This fact defines three relevant axes systems (see Fig.~\ref{fig:S1}): the lab axis, along which we control $B_z^{lab}$ and $B_x^{lab}$ via our magnets; the SQUID axis, with $B_\parallel$ parallel to the SQUID plane and B$_\perp$ perpendicular to it, and finally $B_x,B_z$ which are parallel and perpendicular to the plane of the MLG, respectively. 
When applying magnetic field $B^{lab}_x$ in the lab frame, stray perpendicular flux can penetrate the SQUID, either due to a small misalignment between the sample plane and the magnet axis, or from vortices or trapped magnetic flux in the leads, junction or the magnet itself. This leads to instability in the interference patterns, faster decay in the critical current with $B^{lab}_x$, and difficulty in interpretation of the results. We thus wish to apply high $B_\parallel$ precisely along the SQUID plane, using the alignment procedure described below.

We calibrate $B=0T$ in all directions and axes systems as zero applied field at cooldown. We then apply small $B_z^{lab}$ and measure the interference pattern as a reference: at B$_x^{lab}$=0T, the interference pattern of the supercurrent has a clear maximum at $B_z^{lab}$=0T. Upon applying a small $B_x^{lab}$ (say 10 mT) and subsequently measuring the interference pattern generated by $B_z^{lab}$, there will be a shift in the center of the pattern: since the center of the pattern is located at $B_\perp=0T$, which no longer coincides with $B_z^{lab}=0T$.
To find the true state of zero flux through the SQUID, we assume a smooth evolution of the interference pattern with B$^{lab}_x$; that is, small changes in $B_x^{lab}$ will cause a minimal change/shift in the interference pattern of the junction. Based on this assumption, we find the shift in $B^{lab}_z$ which maximizes the cross-correlation (implemented using Matlab xcorr function) between the measurement at $B_x^{lab}=10mT$ and the reference at $B^{lab}_x$=0T. This shift is the compensation field $B_z^{comp}$, which we then take to be true $B_\perp$=0T. The pattern at 10 mT then serves as a reference for calculating the compensation field at 20 mT and so on. Fig.\ref{fig:S1}b shows the resulting compensation field as a function of $B_x^{lab}$ for the measurements in Fig.~\ref{fig:S3}c and Fig. 4 of the main text. The compensation algorithm along with the entire measurement is automated.

The angle $\theta'\approx0.8\degree$ between the z magnet axis and the normal to the SQUID plane is calculated from the ratio of the compensation field to $B^{lab}_x$. The angle between the MLG plane and the SQUID plane $\theta=0.025\degree$ is then inferred from a fit to the interference pattern as described in Supplementary Section 2. In the main text, as well as what follows in the supplementary, we approximate $B_\parallel=\sqrt{(B_z^{comp})^2+(B_x^{lab})^2}$, and $B_{\perp}=B_z^{lab}-B_z^{comp}$. These approximations are accurate at least until second order in the small parameters $\theta,\theta',\frac{B_\perp}{B_\parallel}$.

The downside of this complicated alignment procedure is that it can also compensate for shifts in the interference pattern due to relevant physical causes, such as changes in the ground phase of the junction leading to a shift in the phase of the interference pattern (a 0-$\pi$ transition for example). This information will not be apparent in a plot of $I_C$ vs $B_\perp,B_\parallel$, such as the one presented in Fig. 4 of the main text. However the information is not lost, since we track the applied compensation field as a function of $B_x^{lab}$ (see Fig.~\ref{fig:S1}b). Note that there was trapped flux in the system in one of the measurements at $B_x^{lab}=0T$, as evidenced by the finite compensation field. Note also that as $B_x^{lab}$ approaches 4T, the compensation curve becomes non-linear; this is due to the automated cross-correlation algorithm gradually failing as the critical current descends below the minimal current detectable by the set voltage threshold. 

\section{Supplementary Section: Analytical calculation of two channel interference pattern}

The Josephson effect occurs when current flows between two superconducting electrodes (in this case NbSe$_2$) connected by a weak link. The proximity of the superconductors, under the correct conditions, allows a supercurrent to flow through the junction. 
Upon application of magnetic field perpendicular to the junction, the superconducting order parameter $\Delta e^{i\varphi}$ acquires a position-dependent phase and undergoes interference, resulting in a diffraction pattern of the critical current in magnetic field.  The first Josephson relation relates the critical current density $J(x)$ to the phase difference between the order parameters of the two superconductors A and B, $\gamma(x)=\varphi_B(x)-\varphi_A(x)$:

\begin{equation}
    J(x)=J_0(x)sin(\gamma(x))
\end{equation}
 
Here $J_0$ is the maximal possible critical current density at location $x$. The order parameter must retain a single-valued phase around any closed loop through which current may circulate, leading to the requirement:

\begin{equation}
    \gamma(x_2)-\gamma(x_1)=\frac{2\pi\phi_A}{\phi_0}
\end{equation}

\begin{figure}[ht!]
    \centering
    \includegraphics[width=0.9\textwidth]{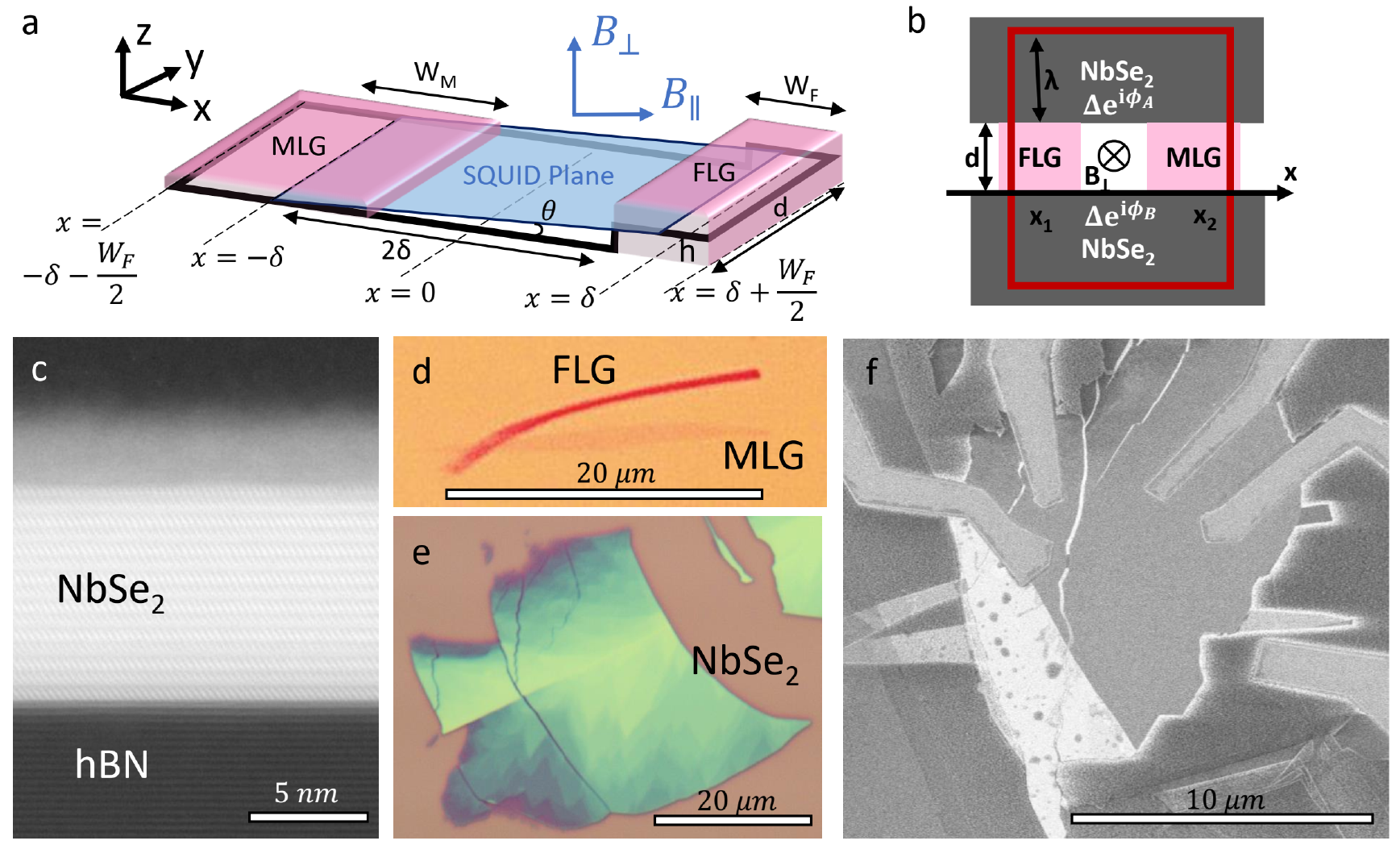}
    \caption{   ~\textbf{a.} Schematic illustration of FLG and MLG parallel weak links of different thicknesses and widths $W_F$,$W_M$ respectively. Directions $x$ and $y$ are in the plane of the flakes, $z$ is perpendicular. Mean SQUID plane is shown in blue, at an angle tan($\theta)=h/2\delta$. $B_\parallel$ is parallel to the SQUID plane and $B_\perp$ is perpendicular to it.  \textbf{b.} Schematic showing B$_\perp$ flux through one possible current circulation path, with an area $(2\lambda+d)|x_2-x_1|$. Crack length $d$ is in the direction of current flow. \textbf{c.} Cross-section TEM measurement of NbSe$_2$ on hBN. The top 4 nm of NbSe$_2$ are oxidized (amorphous, gray color), underneath are 15 visible layers \textbf{d.-e.} Enhanced contrast optical microscope images of graphene and NbSe$_2$ flakes, respectively, exfoliated on SiO$_2$ \textbf{f.} SEM image of the device.}  \label{fig:S2}
\end{figure}

Where $\phi_A$ is the magnetic flux through a loop connecting $x_1,x_2$ and extending across the junction length $d$ into the superconductors up to the London penetration depth $\lambda$ on either side (see Fig.~\ref{fig:S2}b). For convenience we denote $L\equiv 2\lambda +d$.


Now we address our specific geometry. For this device we use NbSe$_2$ of thickness around 12-20 layers ($\approx$7-13 nm). Fifteen layers are clearly seen in cross-section TEM measurement (Fig.~\ref{fig:S2}c), below 3-5 nm of oxide; also from optical images, NbSe$_2$ thickness may vary slightly throughout the sample (Fig.~\ref{fig:S2}e). The junction, of length $d$=140 nm in the direction of the current flow ($y$ axis), has an MLG weak link of width $W_{MLG}\approx 1.45 \mu m$ and an FLG weak link of width $W_{FLG}\approx 0.45 \mu m$ and thickness $h$=2.4 nm (8 layers), their centers separated by a distance $2\delta\approx2.7 \mu m$. We define the $x$ axis in the plane of the MLG/FLG flakes, perpendicular to current flow, and $z$ perpendicular to the MLG/FLG plane. The magnetic field $B_\parallel$ we refer to as ``in-plane" is oriented parallel to the mean SQUID plane: the plane connecting the center of the MLG and FLG flakes. This plane is at a small angle $\tan{(\theta)}=\frac{h}{2\delta}$ with respect to the $x$ axis. The field referred to as $B_\perp$ is perpendicular to $B_\parallel$. This choice in alignment of $B_\parallel$ maintains the peak of the central SQUID oscillation at $B_\perp=0$ regardless of the applied $B_\parallel$  (as described in Section S1). The applied field in $x$,$z$ coordinates is as follows:

\begin{equation}
    B_x=B_\parallel \cos(\theta)-B_\perp \sin(\theta) \approx B_\parallel ,
    B_z=B_\parallel \sin(\theta)+B_\perp \cos(\theta)\approx B_\parallel \sin(\theta)+B_\perp
\end{equation}

The approximations are to the first order in the small parameters $\theta$ and $B_\perp/B_\parallel$. The corresponding wave-numbers are: 

\begin{equation}
    k_1=\frac{2\pi B_x (2\lambda+d)}{\phi_0} ,
    k_2=\frac{2\pi B_z (2\lambda+d)}{\phi_0}
\end{equation}

We define a reference phase $\gamma_0=\gamma(x=0)$, with $x=0$ at the center of the junction, such that the centers of the MLG and FLG flake are both at the same distance $\delta$ from zero.  Finally, the critical current is given by integrating over the current density $J_0(x)sin(\gamma(x))$. We take care that the accumulated phase difference around a closed loop is always 0, accounting for the $B_x$ flux exiting through loops with a vertical portion formed by the step $h$ between the MLG and FlG:

\begin{equation}
    I_C(B_\perp)=\max_{\gamma_0} \bigg{(} \int_{-\infty}^{\infty} J_{M}(x) \sin{(\gamma_0+k_2 x)} + J_{F}(x) \sin{(\gamma_0+k_2 x-k_1 h)} dx \bigg{)}
\end{equation}

The integral can be written as the imaginary part of a complex exponential, and the whole expression becomes a Fourier transform:

\begin{equation}
    I_C(B_\perp)=\max_{\gamma_0} \text{Im} \bigg{(} e^{i\gamma _0} \int_{-\infty}^{\infty} (J_{M}(x) + J_{F}(x) e^{-i k_1 h}) e^{i k_2 x} dx \bigg{)} = \bigg{|}\mathcal{F}\big{(}J_{M}(x) + J_{F}(x) e^{-i k_1 h}\big{)}\bigg{|}
\end{equation}

To get a simple ``zero-order" analytical expression for the critical current in our junction, we normalize the interference pattern by dividing by $I_C(B_\perp=0)$, and assume a constant current density in each channel, with the ratio $\frac{J_F}{J_M}\equiv f$ an unknown parameter. Normalization of $I_C(B_\perp=0)$ implies that $J_F W_F +J_M W_M =1$, and leaves us with the following current densities expressed in terms of $f,W_F,W_M$:

\begin{equation}
    J_{F}=1/(W_{F}+1/f*W_{M});
    J_{M}=1/f*J_{F};
\end{equation}

Thus we obtain: 

\begin{equation}
    I_C(B_\perp)=\bigg{|}\mathcal{F}\bigg{(} J_M \text{rect}\bigg{(}\frac{x+\delta}{W_M}\bigg{)} + J_{F} \text{rect}\bigg{(}\frac{x-\delta}{W_F}\bigg{)}  e^{-i k_1 h}\bigg{)}\bigg{|}
\end{equation}

And the analytical expression for the interference pattern:

\begin{equation}
    \begin{aligned}
    \frac{I_C(B_\perp)}{I_C(B_\perp=0)} & =\sqrt{I_F^2+I_M^2+2*I_F*I_M*\cos{(k_2*2\delta+k_1 h)}} \\
    I_F & \equiv J_F W_F *\sinc{(k_2 W_F)}\\
    I_M & \equiv J_M W_M *\sinc{(k_2 W_M)}
    \end{aligned}
\end{equation}

The angle of the SQUID plane with respect to the MLG and FLG plane translates into a phase difference between the two channels. Following is a tabulation of the fit parameters $W_M,W_F,2\delta,\theta,h$ and their errors, extracted from a Matlab non-linear least squares fit of data shown in main text Figs. 3, 4 to the analytical model. The extracted dimensions may be compared to the measured dimensions given above. Note that there are different combinations of slightly different parameter values that can also yield a similar fit, therefore the true fit error is larger than the error bars given in the table.

\begin{center}
\begin{tabular}{||c c c c c c c||} 
 \hline
 $V_g$ & $B_\parallel$ & $W_F$ & $W_M$ & $2\delta$ & $\theta$ & h \\ [0.5ex] 
 \hline\hline
 0 V & 0 T & 330 $\pm$ 10 nm & 1260 $\pm$ 25 nm & 2720 $\pm$ 15 nm & X & X  \\ 
 \hline
  -30 V & 0 T & 310 $\pm$ 20 nm & 1590 $\pm$ 40 nm & 2740 $\pm$ 20 nm & X & X \\
 \hline
 30 V & 0 T & 300 $\pm$ 15 nm & 1600 $\pm$ 55 nm & 2620 $\pm$ 30 nm & X & X \\
 \hline
   0 V & 0 T & 310 $\pm$ 50 nm & 1500 $\pm$ 60 nm & 2600 $\pm$ 35 nm & X & X \\ 
   \hline
  0 V & 2.49 T & 210 $\pm$ 20 nm & 315 $\pm$ 40 nm & 2500 $\pm$ 15 nm & 0.024 $\pm0.0003\degree$ & 1.1 $\pm$ 0.01 nm \\
 [1ex] 
 \hline
\end{tabular}
\end{center}

\section{Supplementary Section: Maximum entropy reconstruction of the current profile via Markov Chain Monte Carlo simulated annealing}

In order to extract the current distribution in greater detail, we postulate an initial current density profile sampled at N discrete spatial points and subject to physical constraints, and calculate the corresponding interference pattern. We then adjust the density profile sequentially to obtain the best least-squares fit of the calculated interference pattern to the data, subject to a maximum entropy constraint. This is done via Markov chain Monte Carlo simulated annealing. 

We begin by guessing a current density vector sampled at N points $\bar{J}_0 = [J_0(x_1),J_0(x_2)...J_0(x_N)]$ normalized such that $\sum_n{J_x(x_n)} = 1.$ Valid guesses are constrained such that non-zero current density can exist only within the MLG and FLG channels, and current reversal (negative $\bar{J}_0$) is disallowed. We then calculate the corresponding interference pattern based on equation S9, with the perpendicular field sampled at M points $\bar{B}_\perp=[B_\perp^1,B_\perp^2...B_\perp^M]$, $B_\parallel$ set to some value, step height $h=1$ nm and junction length $L=2.2 \mu m$ extracted from the analytical fit described in Section S2. Explicitly, we define the matrix element $A^m_{n}=\exp{(i\frac{2\pi}{\phi_0} L B_z^m x_n)}$ for $x_n<0$, and $A^m_{n}=\exp{(i\frac{2\pi}{\phi_0} L ( B_z^m x_n-B_x^m h))}$ for $x_n>0$. We then compute:

\begin{equation}
    I_C^{calc}(B^m_\perp)= \bigg{|}\sum_n A^m_nJ_0(x_n)\bigg{|}
\end{equation}

In order to quantify the fit of our guess $\bar{J}_0$ we calculate the least squares difference of $I_C^{calc}(B^m_\perp)$ with respect to the measured interference pattern $I_C^{meas}(B^m_\perp)$:

\begin{equation}
    \chi^2 (\bar{J}_0)=\sum^M_{m=1}(I_C^{calc}(B^m_\perp)-I_C^{meas}(B^m_\perp))^2
\end{equation}

We then sequentially adjust the fit by employing a Metropolis algorithm Markov chain Monte Carlo (MCMC) process which samples possible $\bar{J}_0$ configurations and assigns them a free energy reflecting a competition between the goodness of fit $\chi^2$ and the entropy.

\begin{equation}
    F(\bar{J}_0)=\chi^2(\bar{J}_0)+\lambda\sum^N_{n=1}J_0(x_n)\ln{(J_0(x_n))}
\end{equation}

Samples are correspondingly weighted with the standard Boltzmann weight $e^{-\beta F}$. The first term in the free energy penalizes a large deviation of the fit from the measurement, the second (entropy) term penalizes non-uniformity of the postulated current distribution, and the hyper-parameter $\lambda$ tunes between them. A finite ``temperature" $T=\beta^{-1}>0$ introduces noise to the equilibrium current distribution but allows the algorithm to consider corrections to $\bar{J}_0$ which result in energy loss, and thus helps to avoid converging to local minima of $\chi^2$. In order to find the minimum of $F$, we employ simulated annealing, increasing the inverse temperature from 0 to $\beta$ linearly with each Monte Carlo (MC) step. We ensure that $\bar{J}_0$ remains normalized by making changes in  discrete units of size $\Delta J$; a unit removed from site $x_i$ must be added to some other site $x_j$. The steps of the algorithm are as follows: 

\begin{enumerate}
  \item Make an initial guess $\bar{J}_0$ obeying normalization and constraints. 
  \item Calculate $F(\bar{J}_0)$
  \item Choose $x_i,x_j$ at random from among the sites at which $\bar{J}_0$ is allowed to be non-zero. Propose $J'_0(x_i)=J_0(x_i)+\Delta J , J'_0(x_j)=J_0(x_j)-\Delta J$ so that the total current density is conserved. 
  \item If either $J'_0(x_i),J'_0(x_j)$ is negative, return to step 3. Otherwise, continue.
  \item Calculate the new free energy $F_{new}(\bar{J}'_0)$
  \item If $F_{new}<F$, accept new current density and return to step 3 updating $\bar{J}_0 \rightarrow \bar{J}'_0$. Otherwise, accept change $\bar{J}_0 \rightarrow \bar{J}'_0$ with probability $e^{-\beta_i(F_{new}-F)}$. The temperature at step $i$ is given by $\beta_i=\beta/K*i$, where $K$ is the total number of MC steps. 
  \item Return to step 3 and iterate K times.
\end{enumerate}

We have freedom to change our initial guess $\bar{J}_0$, and to tune the hyper-parameters $\lambda,\beta,N,\Delta J,K$.
The $x$ range of the simulation is defined between $\pm2 \mu m$, while the current in the MLG/FLG channels is bounded within the widths determined from the SEM measurement. Fourier uncertainty indicates that an interference pattern with $N$ nodes yields a spatial resolution of $\frac{W}{N}$; that is, we can choose N evenly spaced discrete points to sample within the overall width $W$ of the current carrying channels. However our fitting method is not an inverse Fourier transform; it introduces additional information through geometrical constraints as well as the maximum entropy constraint. Increasing the number of sampling points helps better fit the width and separation of the channels, while the maximum entropy constraint smooths any sharp spatial features. Thus, we choose $N=50$, a few times larger than the bandwidth. We use a uniform initial distribution $\bar{J}_0=\frac{1}{N^*}$, with $N^*$ being the number of sample points allowed to carry current.


\begin{figure}[ht!]
    \centering
    \includegraphics[width=0.9\textwidth]{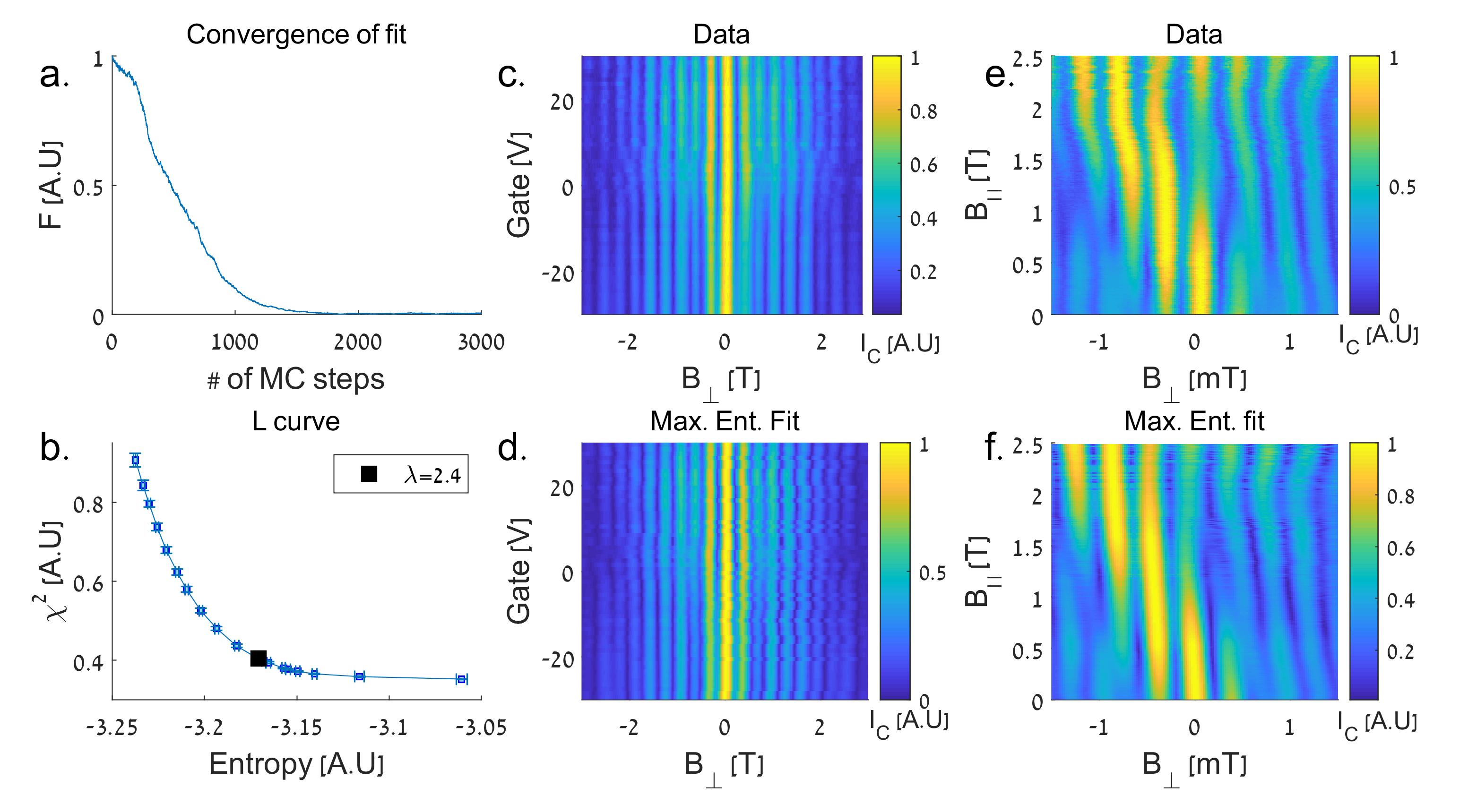}
    \caption{   ~\textbf{a.} Convergence of the free energy as a function of MC steps ~\textbf{b.} The L curve, a parametric plot of $\chi^2$ and entropy for different values of $\lambda$, with chosen $\lambda\approx2.4$ marked ~\textbf{c.} Normalized $I_C$ (color scale) vs. $B_\perp$ and $V_G$ ~\textbf{d.} Normalized $I_C$ (color scale) vs. $B_\perp$ and gate calculated using Eq. S6, with current density extracted from panel (a) by the Monte Carlo method~\textbf{e.} Normalized $I_C$ (color scale) vs. $B_\perp$ and $B_\parallel$ ~\textbf{f.} Normalized $I_C$ (color scale) vs. $B_\perp$ and $B_\parallel$ from eq. S6, with current density extracted from panel (c) by the Monte Carlo method } \label{fig:S3}
\end{figure}


At zero magnetic field and maximal charge carrier density ($V_G$=-1V) we have measured a stable, nearly symmetric interference pattern with many lobes (see main text Fig. 3). We use this pattern to tune all of the hyper-parameters of the fitting algorithm. To tune the parameters $\beta, \Delta J$ we set $\lambda=0$ and try several initial guesses for $\beta, \Delta J$ in powers of 10 before settling on  $\beta=10000,\Delta J=0.001$ to obtain a convergence of $F$ in a few thousand MC steps (see \ref{fig:S3}a). We note that any $10^3<\beta$ and $\Delta J<10^{-2}$ would work as well. We then choose $K=5000$, several thousand MC steps after convergence. Finally, we tune $\lambda$ by studying the L curve (see \ref{fig:S3}b), and choosing $\lambda=2.4$ which provides a trade-off between decrease in goodness of fit and increase in entropy. Any $\lambda$ in the vicinity (up to $\approx 6$) gives a qualitatively similar current distribution, with smaller $\lambda$ generating a noisier distribution and larger $\lambda$ providing a poorer fit.  As a sanity check, the spatial features of the current distribution generated by the fit do not have significant features on a length scale finer than that offered by the bandwidth of the original signal, as can be seen in Figs. 3, 4 of the main text. These parameters, chosen once based on the best data set, were then used to fit all of our measured data. See for example measured data of $I_C$ vs. $B_\perp$ and gate voltage (\ref{fig:S3}c) and $I_C$ vs. $B_\perp$,$B_\parallel$) (\ref{fig:S3}e), and compare to the fit in panels d,f. We note that there can in principle be multiple current distributions which give a comparable fit to the interference pattern, due to the loss of phase information. However, fitting three different experimental repetitions of the $B_\parallel$ measurement, changing the fitting hyper-parameters, changing initial conditions, etc. in the above described fitting procedure always results in similar current distributions if we use $\lambda$ in the vicinity of the optimal $\lambda$ determined by the L curve.

\section{Supplementary section: Current density extraction using the Wiener Khinchin theorem}

The Wiener Khinchin theorem states that the energy spectral density of a function and its autocorrelation $C(l)$ are Fourier transform pairs. In our case, $|I_C(B_\perp)|^2$ is the energy spectral density of $J_0(x)$, and thus:

\begin{equation}
        F(|I_C(B_\perp)|^2)=C(l)=\int_{-\infty}^{\infty}J_0^*(x) J_0(x+l) dx
\end{equation}

Consider a two channel device, with the current density in the first channel very sharp and narrow, approximated by the Dirac delta function located at $x=a$, and the current density in the second channel given by some function $F$ of finite width W centered at $x=-a$. The current density is thus: J$_0$(x)=$\delta(x-a)+F(x+a)$. The autocorrelation of the current density in this case is:

\begin{equation}
        C(l)=\int_{-\infty}^{\infty}\delta(x-a)\delta(x-a+l)+F(x+a)F(x+a+l)+F(x+a)\delta(x-a+l)+F(x+a+l)\delta(x-a) dx
\end{equation}

The first two terms give some function centered at $l=0$. If we assume W$<$a, this function extends no further than $l=\pm W$. The second two terms give $F(2a \pm l)$: this is the current density of the second channel, centered at shifts equal to the distance between the two channels, $l=\pm 2a$ (and mirrored with respect to $l$ around $l=2a$). In our case the FLG is only a few times narrower than the MLG, and carries a similar current density. In this instance, the autocorrelation convolves the FLG and MLG densities, resulting in a feature which qualitatively resembles the MLG current density ``smeared" at the scale of the FLG width, and centered at $l=-2\delta$ equal to the distance between the centers of the two channels.


\begin{thebibliography}{10}

\bibitem{Xi_2016}
Xiaoxiang Xi, Zefang Wang, Weiwei Zhao, Ju-Hyun Park, Kam~Tuen Law, Helmuth
  Berger, L{\'a}szl{\'o} Forr{\'o}, Jie Shan, and Kin~Fai Mak.
\newblock {Ising pairing in superconducting NbSe$_2$ atomic layers}.
\newblock {\em Nature Physics}, 12(2):139--143, February 2016.

\bibitem{Tsen2016}
A.~W. Tsen, B.~Hunt, Y.~D. Kim, Z.~J. Yuan, S.~Jia, R.~J. Cava, J.~Hone,
  P.~Kim, C.~R. Dean, and A.~N. Pasupathy.
\newblock Nature of the quantum metal in a two-dimensional crystalline
  superconductor.
\newblock {\em Nature Physics}, 12(3):208--212, Mar 2016.

\bibitem{Dvir_Nat_Comm_2018}
T~Dvir, F~Massee, L~Attias, M~Khodas, Marco Aprili, Charis H~L Quay, and Hadar
  Steinberg.
\newblock {Spectroscopy of bulk and few-layer superconducting NbSe$_2$ with van
  der Waals tunnel junctions}.
\newblock {\em Nature Communications}, 9:598, 2018.

\bibitem{DvirArxiv}
M.~Kuzmanović, T.~Dvir, D.~LeBoeuf, S.~Ilić, M.~Haim, D.~Möckli, S.~Kraemer,
  M.~Khodas, M.~Houzet, J.~S. Meyer, M.~Aprili, H.~Steinberg, and C.~H.~L.
  Quay.
\newblock {Tunneling spectroscopy of few-monolayer NbSe$_2$ in high magnetic
  field: Ising protection and triplet superconductivity}.
\newblock {\em arXiv}, arXiv:2104.00328, 2021.

\bibitem{Efetov_2016}
D~K Efetov, L~Wang, C~Handschin, K~B Efetov, J~Shuang, R~Cava, T~Taniguchi,
  K~Watanabe, J~Hone, C~R Dean, and P~Kim.
\newblock {Specular interband Andreev reflections at van der Waals interfaces
  between graphene and NbSe$_2$}.
\newblock {\em Nature Physics}, 12(4):328--332, April 2016.

\bibitem{Yabuki2020}
Rai Moriya, Naoto Yabuki, and Tomoki Machida.
\newblock Superconducting proximity effect in a
  $\mathrm{Nb}{\mathrm{Se}}_{2}/\text{graphene}$ van der Waals junction.
\newblock {\em Phys. Rev. B}, 101:054503, Feb 2020.

\bibitem{Lee2019}
Jongyun Lee, Minsoo Kim, Kenji Watanabe, Takashi Taniguchi, Gil~Ho Lee, and
  Hu~Jong Lee.
\newblock {Planar graphene Josephson coupling via van der Waals superconducting
  contacts}.
\newblock {\em Current Applied Physics}, 19(3):251--255, 2019.

\bibitem{Zalic_2021}
Tom Dvir, Ayelet Zalic, Eirik~Holm Fyhn, Morten Amundsen, Takashi Taniguchi,
  Kenji Watanabe, Jacob Linder, and Hadar Steinberg.
\newblock Planar graphene-${\mathrm{Nbse}}_{2}$ Josephson junctions in a
  parallel magnetic field.
\newblock {\em Phys. Rev. B}, 103:115401, Mar 2021.

\bibitem{Tinkham}
M.~Tinkham.
\newblock {\em Introduction to Superconductivity}.
\newblock International series in pure and applied physics. McGraw Hill, 1996.

\bibitem{Allen2016}
M~T Allen, O~Shtanko, I~C Fulga, A~R Akhmerov, K~Watanabe, T~Taniguchi,
  P.~Jarillo-Herrero, L~S Levitov, and A~Yacoby.
\newblock {Spatially resolved edge currents and guided-wave electronic states
  in graphene}.
\newblock {\em Nature Physics}, 12(2):128--133, feb 2016.

\bibitem{Zhu2017}
M.~J. Zhu, A.~V. Kretinin, M.~D. Thompson, D.~A. Bandurin, S.~Hu, G.~L. Yu,
  J.~Birkbeck, A.~Mishchenko, I.~J. Vera-Marun, K.~Watanabe, T.~Taniguchi,
  M.~Polini, J.~R. Prance, K.~S. Novoselov, A.~K. Geim, and M.~Ben Shalom.
\newblock {Edge currents shunt the insulating bulk in gapped graphene}.
\newblock {\em Nature Communications}, 8:6--11, 2017.

\bibitem{Nanda2017}
G.~Nanda, J.~L. Aguilera-Servin, P.~Rakyta, A.~Korm{\'a}nyos, R.~Kleiner,
  D.~Koelle, K.~Watanabe, T.~Taniguchi, L.~M.~K. Vandersypen, and S.~Goswami.
\newblock Current-phase relation of ballistic graphene Josephson junctions.
\newblock {\em Nano Letters}, 17(6):3396--3401, Jun 2017.

\bibitem{Rodan-Legrain2021}
Daniel Rodan-Legrain, Yuan Cao, Jeong~Min Park, Sergio~C. de~la Barrera,
  Mallika~T. Randeria, Kenji Watanabe, Takashi Taniguchi, and Pablo
  Jarillo-Herrero.
\newblock Highly tunable junctions and non-local Josephson effect in
  magic-angle graphene tunnelling devices.
\newblock {\em Nature Nanotechnology}, 16(7):769--775, Jul 2021.

\bibitem{Jaklevic1965}
R.~C. Jaklevic, J.~Lambe, J.~E. Mercereau, and A.~H. Silver.
\newblock Macroscopic quantum interference in superconductors.
\newblock {\em Phys. Rev.}, 140:A1628--A1637, Nov 1965.

\bibitem{Rasmussen_2016}
Asbj\o{}rn Rasmussen, Jeroen Danon, Henri Suominen, Fabrizio Nichele, Morten
  Kjaergaard, and Karsten Flensberg.
\newblock Effects of spin-orbit coupling and spatial symmetries on the
  Josephson current in SNS junctions.
\newblock {\em Phys. Rev. B}, 93:155406, Apr 2016.

\bibitem{Assouline_2019}
Alexandre Assouline, Cheryl Feuillet-Palma, Nicolas Bergeal, Tianzhen Zhang,
  Alireza Mottaghizadeh, Alexandre Zimmers, Emmanuel Lhuillier, Mahmoud Eddrie,
  Paola Atkinson, Marco Aprili, and Herv{\'e} Aubin.
\newblock {Spin-Orbit induced phase-shift in Bi$_2$Se$_3$ Josephson junctions}.
\newblock {\em Nature Communications}, 10(1):126, January 2019.

\bibitem{Dynes_1971}
R.~C. Dynes and T.~A. Fulton.
\newblock {Supercurrent Density Distribution in Josephson Junctions}.
\newblock {\em Phys. Rev. B}, 3:3015--3023, May 1971.

\bibitem{Hui_2014}
Hoi-Yin Hui, Alejandro~M. Lobos, Jay~D. Sau, and S.~Das~Sarma.
\newblock {Proximity-induced superconductivity and Josephson critical current
  in quantum spin Hall systems}.
\newblock {\em Phys. Rev. B}, 90:224517, Dec 2014.

\bibitem{Ghatak2018}
Subhamoy Ghatak, Oliver Breunig, Fan Yang, Zhiwei Wang, Alexey~A. Taskin, and
  Yoichi Ando.
\newblock {Anomalous Fraunhofer Patterns in Gated Josephson Junctions Based on
  the Bulk-Insulating Topological Insulator BiSbTeSe$_2$}.
\newblock {\em Nano Letters}, 18(8):5124--5131, Aug 2018.

\bibitem{Hansen_1992}
Per~Christian Hansen.
\newblock {Analysis of Discrete Ill-Posed Problems by Means of the L-Curve}.
\newblock {\em SIAM Review}, 34:561--80, Dec 1992.

\bibitem{Suominen2017}
H.~J. Suominen, J.~Danon, M.~Kjaergaard, K.~Flensberg, J.~Shabani, C.~J.
  Palmstr\o{}m, F.~Nichele, and C.~M. Marcus.
\newblock {Anomalous Fraunhofer interference in epitaxial
  superconductor-semiconductor Josephson junctions}.
\newblock {\em Phys. Rev. B}, 95:035307, Jan 2017.

\bibitem{Buzdin2005}
A.~I. Buzdin.
\newblock Proximity effects in superconductor-ferromagnet heterostructures.
\newblock {\em Rev. Mod. Phys.}, 77:935--976, Sep 2005.

\bibitem{Pannetier2000}
B.~Pannetier and H.~Courtois.
\newblock Andreev reflection and proximity effect.
\newblock {\em Journal of Low Temperature Physics}, 118(5):599--615, Mar 2000.

\bibitem{Gmitra2015}
Martin Gmitra and Jaroslav Fabian.
\newblock Graphene on transition-metal dichalcogenides: A platform for
  proximity spin-orbit physics and optospintronics.
\newblock {\em Phys. Rev. B}, 92:155403, Oct 2015.

\bibitem{Li_2019}
Chuan Li, Bob de~Ronde, Jorrit de~Boer, Joost Ridderbos, Floris Zwanenburg,
  Yingkai Huang, Alexander Golubov, and Alexander Brinkman.
\newblock {Zeeman-Effect-Induced 0-$\pi$ Transitions in Ballistic Dirac
  Semimetal Josephson Junctions}.
\newblock {\em Physical Review Letters}, 123(2):026802, 2019.

\bibitem{Ke_2019}
Chung~Ting Ke, Christian~M Moehle, Folkert~K Vries, Candice Thomas, Sara Metti,
  Charles~R Guinn, Ray Kallaher, Mario Lodari, Giordano Scappucci, Tiantian
  Wang, Rosa~E Diaz, Geoffrey~C Gardner, Michael~J Manfra, and Srijit Goswami.
\newblock {Ballistic Superconductivity and Tunable $\pi$ Junctions in InSb
  Quantum Wells}.
\newblock {\em Nature Communications}, 10(1):3764, August 2019.

\bibitem{Hart_2016}
Sean Hart, Hechen Ren, Michael Kosowsky, Gilad Ben-Shach, Philipp Leubner,
  Christoph Br{\"u}ne, Hartmut Buhmann, Laurens~W Molenkamp, Bertrand~I
  Halperin, and Amir Yacoby.
\newblock {Controlled finite momentum pairing and spatially varying order
  parameter in proximitized HgTe quantum wells}.
\newblock {\em Nature Physics}, 13(1):87--93, September 2016.

\bibitem{Chen_2018}
Angela~Q Chen, Moon~Jip Park, Stephen~T Gill, Yiran Xiao, Dalmau Reig-i
  Plessis, Gregory~J MacDougall, Matthew~J Gilbert, and Nadya Mason.
\newblock {Finite momentum Cooper pairing in three-dimensional topological
  insulator Josephson junctions}.
\newblock {\em Nature Communications}, 9(1):3478, August 2018.

\bibitem{Pientka_2017}
Falko Pientka, Anna Keselman, Erez Berg, Amir Yacoby, Ady Stern, and Bertrand~I
  Halperin.
\newblock {Topological Superconductivity in a Planar Josephson Junction}.
\newblock {\em Physical Review X}, 7(2):021032, May 2017.

\bibitem{Ren_2019}
Hechen Ren, Falko Pientka, Sean Hart, Andrew~T Pierce, Michael Kosowsky, Lukas
  Lunczer, Raimund Schlereth, Benedikt Scharf, Ewelina~M Hankiewicz, Laurens~W
  Molenkamp, Bertrand~I Halperin, and Amir Yacoby.
\newblock {Topological superconductivity in a phase-controlled Josephson
  junction}.
\newblock {\em Nature}, 569(7754):93--98, April 2019.

\bibitem{Fornieri_2019}
Antonio Fornieri, Alexander~M Whiticar, F~Setiawan, El{\'\i}as Portol{\'e}s,
  Asbj{\o}rn C~C Drachmann, Anna Keselman, Sergei Gronin, Candice Thomas, Tian
  Wang, Ray Kallaher, Geoffrey~C Gardner, Erez Berg, Michael~J Manfra, Ady
  Stern, Charles~M Marcus, and Fabrizio Nichele.
\newblock {Evidence of topological superconductivity in planar Josephson
  junctions}.
\newblock {\em Nature}, 569(7754):89--92, May 2019.

\bibitem{Mayer2019B}
William {Mayer}, Matthieu~C. {Dartiailh}, Joseph {Yuan}, Kaushini~S.
  {Wickramasinghe}, Enrico {Rossi}, and Javad {Shabani}.
\newblock {Gate controlled anomalous phase shift in Al/InAs Josephson
  junctions}.
\newblock {\em Nature Communications}, 11:212, January 2020.

\bibitem{Island2019}
J.~O. Island, X.~Cui, C.~Lewandowski, J.~Y. Khoo, E.~M. Spanton, H.~Zhou,
  D.~Rhodes, J.~C. Hone, T.~Taniguchi, K.~Watanabe, L.~S. Levitov, M.~P.
  Zaletel, and A.~F. Young.
\newblock {Spin-orbit-driven band inversion in bilayer graphene by the van der
  Waals proximity effect}.
\newblock {\em Nature}, 571(7763):85--89, Jul 2019.

\bibitem{Gani2019}
Yohanes~S. Gani, Hadar Steinberg, and Enrico Rossi.
\newblock Superconductivity in twisted graphene ${\mathrm{NbSe}}_{2}$
  heterostructures.
\newblock {\em Phys. Rev. B}, 99:235404, Jun 2019.

\bibitem{Zomer2014}
P.~J. Zomer, M.~H.~D. Guimarães, J.~C. Brant, N.~Tombros, and B.~J. van Wees.
\newblock Fast pick up technique for high quality heterostructures of bilayer
  graphene and hexagonal boron nitride.
\newblock {\em Applied Physics Letters}, 105(1):013101, 2014.

\end{thebibliography}
\end{document}